\documentclass{llncs}
\usepackage{multirow}
\usepackage{mathrsfs}
\usepackage{amsmath}
\usepackage{amssymb}
\usepackage{array}
\usepackage[bookmarks=false]{hyperref}
\usepackage{graphicx}
\usepackage{epstopdf}
\usepackage{multicol}
\usepackage{xcolor}
\usepackage{subcaption}

\usepackage{soul}
\usepackage[bordercolor=white,backgroundcolor=orange!30,linecolor=black,colorinlistoftodos]{todonotes}

\begin{document}

\title{Lexical-Morphological Modeling for\\ Legal Text Analysis}
\titlerunning{Lexical-Morphological Model for Legal Text}  % abbreviated title (for running head)
%                                     also used for the TOC unless
%                                     \toctitle is used
%
\author{Danilo S. Carvalho\inst{1}\thanks{Supported by CNPq -- Brazil scholarship grant} \and Minh-Tien Nguyen\inst{1,2} \and Chien-Xuan Tran\inst{1} \and \\ Minh-Le Nguyen\inst{1}}
\authorrunning{Carvalho et al.} % abbreviated author list (for running head)
%
%%%% list of authors for the TOC (use if author list has to be modified)
\tocauthor{Danilo S. Carvalho, Minh-Tien Nguyen, Tran Xuan Chien and Minh Le Nguyen}
\institute{School of Information Science,\\ Japan Advanced Institute of Science and Technology (JAIST),\\ 
1-1 Asahidai, Nomi, Ishikawa, 923-1292, Japan.
\and 
Hung Yen University of Education and Technology (UTEHY), Hung Yen, Vietnam. \\
\email{{\{danilo, tiennm, chien-tran, nguyenml\}}@jaist.ac.jp}}

\maketitle              % typeset the title of the contribution

\begin{abstract}
 In the context of the Competition on Legal Information Extraction/Entailment (COLIEE), we propose a method comprising the necessary steps for finding relevant documents to a legal question and deciding on textual entailment evidence to provide a correct answer. The proposed method is based on the combination of several lexical and morphological characteristics, to build a language model and a set of features for Machine Learning algorithms. We provide a detailed study on the proposed method performance and failure cases, indicating that it is competitive with state-of-the-art approaches on Legal Information Retrieval and Question Answering, while not needing extensive training data nor depending on expert produced knowledge. The proposed method achieved significant results in the competition, indicating a substantial level of adequacy for the tasks addressed.
\end{abstract}
\section{Introduction} % Danilo
\vspace{-0.2cm}
Answering legal questions has been a long-standing challenge in the Information Systems research landscape. This topic draws progressively more attention, as we experience an explosive growth in legal document availability on the World Wide Web and specialized systems. This growth is not accompanied by a matching increase in information analysis capabilities, which points to a severe under-utilization of available resources and to potential for information quality issues~\cite{R08}. As a consequence, increasing pressure has been put into professionals of law, since having the relevant and correct information is a vital step in legal case solving and thus is closely tied to the matter of professional ethics and liability. This problem is often referred as the ``information crisis'' of law.

The ability to retrieve relevant and correct information given a legal query has improved over time, with the combination of expert Knowledge Engineering and Natural Language Processing (NLP) methods. However, the ability to answer questions in the legal domain is of special difficulty, due to the need of reasoning over different types of information, such as past decisions, laws and facts. Furthermore, concepts in legal text are often used in a way that differs from common language use, and differences in laws and procedures from each country prevent the creation of comprehensive and coherent international law corpora. Common legal ontologies are among the efforts to facilitate automatic legal reasoning, but have not seen strong development in the past years~\cite{HBBA07}. In this context, \textit{Textual Entailment Recognition} plays a very important role, as a set of hypothesis presented in a question will certainly have answers in the previously cited types of information (decisions, laws, facts). The Recognition of Textual Entailment (RTE) challenge series~\footnote{\href{http://www.aclweb.org/aclwiki/index.php?title=Recognizing\_Textual\_Entailment}{www.aclweb.org/aclwiki/index.php?title=Recognizing\_Textual\_Entailment}}, although not specific to the legal domain, is a recognized benchmark for methods that can be adapted to legal texts.

    To effectively answer legal questions, one fundamental set of information that must be available is the law, presented as the collection of codes, sections, articles and paragraphs that should be unequivocally referenced when a hypothesis is raised as part of a legal inquiry. Therefore, adequate representation of law corpora is the basis of a functional system for legal question answering. The representation problem is often associated with ontologies and other annotated knowledge bases, but these methods are costly and more difficult to automate when compared to fully text-based approaches, such as bag-of-words, n-gram and topic models.

    In this work, we propose a fully text-based method for legal text analysis, in the context of the Competition on Legal Information Extraction/Entailment (COLIEE), covering both the tasks of Information Extraction and Question Answering. The goal is the retrieval of relevant law articles to a given yes/no legal question and the use of the retrieved articles to correctly answer the question in a completely automated way. Our contributions in this paper are as follows: (i) a ranking and selection method for legal information retrieval based on a mixed size n-gram model, including an original scoring function for ranking; (ii) an improved adaptation of a Textual Entailment classification method, based on Machine Learning ensembles (Adaboost), including a similarity feature built upon Distributional Semantics (Word2Vec). Lexical and morphological analysis were done on the English translated Japanese Civil Code, comprising tokenization, POS-tagging, lemmatization, word clustering and a set of lexical statistics. A study on success and fail cases is also provided, with common baseline practices and related works used as means of performance comparison. The results of COLIEE are presented as a means of substantiating the experimental evaluation and also discussing the proposed method's perceived shortcomings and improvements.  

    The remaining of this work is structured as follows: Section~\ref{sec:related} presents the related works and relevant results; Section~\ref{sec:LQA} details the Legal Question Answering problem and the COLIEE competition shared task; Section~\ref{sec:approach} explains our approach to the competition problem; Section~\ref{sec:experiments} presents the experimental setting, results and discussion; Finally, Section~\ref{sec:conclusion} offers some concluding remarks.

\section{Related Works}\label{sec:related} 
\vspace{-0.3cm}
Liu, Chen and Ho~\cite{LCH15} presented the three-phase prediction (TPP) method for retrieval of relevant statutes in Taiwan's criminal law, given general language queries. The method was a hierarchical ranking approach to law corpora, featuring a combination of several Information Retrieval techniques, as well as Machine Learning and feature selection ones. Results were evaluated in terms of recall, achieving from 0.52 to 0.91, from the top 3 to 10 retrieved results, respectively.

Inkpen et al. showed one of the first successful models for RTE using SVMs \cite{IKN06}. Later, Castillo proposed a new system for solving RTE using SVMs~\cite{C10}, in which training data includes RTE-3, annotated data set from RTE-4, and the development set of RTE-5. 32 features were used and the training model achieved the best F-measure of 0.69 in two-way and 0.67 in three-way classification task.

Nguyen et al. \cite{NHNNN15} conducted a study of RTE on a Vietnamese version of RTE-3 \cite{PNS12} translated from Giampiccolo et. al.~\cite{GMDD07}. The author used SVMs trained with 15 features divided in two groups: distance and statistical features, in which the first group captures the distance and the second one represents the word overlapping between two sentences. A voting system combining three classifiers built on three feature groups (distance, statistical, and combined features) was used to judge entailment relation. The method obtained 0.684 of F-measure in two-way task.

In legal text, Tran et al. addressed legal text QA by using inference \cite{TNNS14}. The author used requisite-effectuation structures of legal sentences and similarity measures to find out correct answers without training data and achieved 60.8\% accuracy on 51 articles on Japanese National Pension Law.

Kim et al. proposed a hybrid method containing simple rules and unsupervised learning using deep linguistic features to address RTE in civil law \cite{KXGS14}. The author also constructed a knowledge base for negation and antonym words which would be used for classifying simple questions. To deal with difficult questions, the author used morphological, syntactic and lexical analysis to identify premises and conclusions. The accuracy was 68.36\% with easy questions and 60.02 with difficult ones.

This work uses all features in \cite{NHNNN15}, as they apply to the same purpose. Additional features were also included: Word2Vec similarity and term frequency -- inverse document frequency (TF-IDF). Our approach differs from \cite{C10} in using \textit{Word2Vec}\cite{MSCCD13} similarity instead of \textit{WordNet}.

\vspace{-0.3cm}

\section{Legal Question Answering}\label{sec:LQA} % Tien
\vspace{-0.3cm}

\begin{figure*}[ht]
	\centering
	\includegraphics[width=0.8\textwidth]{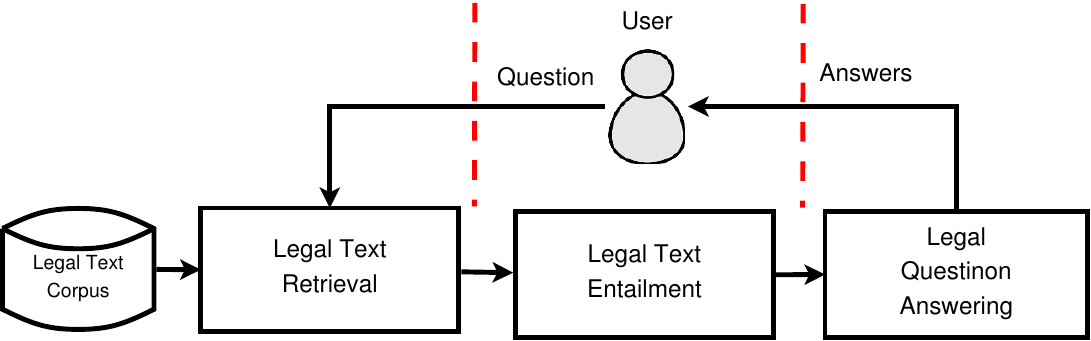}
	\caption{The model of legal text question answering system}\label{fig:model}
\end{figure*}

Legal Question Answering (LQA) consists in finding out and providing \emph{``correct answers"} to a legal question given by users. An overview of LQA is shown in Fig.~\ref{fig:model}.

LQA can be divided in three tasks: 1) retrieving relevant articles, i.e., the ones containing the answer; 2) finding correct evidence in the relevant articles that allows answering the question; and 3) answering the question. While the first task is a specific case of Information Retrieval (IR), the second can be considered as a form of Recognition of Textual Entailment (RTE), in which given a question, the LQA system has to decide whether and how a relevant article can answer the question. The third one is the final result of the two previous tasks, combined with answer formatting.

Legal text is considerably different to other types of text, e.g., news articles, due to their structural and semantic characteristics. Firstly, they have specific logical sentence structures e.g., requisite and effectuation \cite{NNS10-1}. Secondly, words and writing style are used in a strict form, because law documents require high correctness and should avoid ambiguity. Another aspect is that law documents are written in a high abstraction level \cite{TNNS13}; therefore, they often require collection and linking of multiple concept references to enable understanding and answering of a question. The use of concept references leads to a situation in which there are few, or in some cases, even no overlapping words between a law question and its relevant articles.
    
In this work, LQA tasks are considered into the context of COLIEE, a competition on legal information extraction/entailment which was first held in 2014, in association with Workshop on Juris-informatics (JURISIN). COLIEE 2015 \footnote{\href{http://webdocs.cs.ualberta.ca/\~miyoung2/COLIEE2015/}{webdocs.cs.ualberta.ca/\~miyoung2/COLIEE2015/}} is the second competition, consisting of three phases:
\begin{itemize}
	\item Phase One: retrieving relevant articles from all Japanese Civil Code Articles given a set of YES/NO questions.
    \item Phase Two: evaluating the entailment relationship between the question and retrieved articles.
    \item Phase Three: combination of Phase One and Phase Two, the system will retrieve list of relevant articles given a query, and then decide the entailment relationship between retrieved articles and the provided question.
\end{itemize}

The Japanese Civil Code is composed by a collection of numbered articles, each one containing a set of declarations pertaining to a specific topic of the law, e.g., labor contracts, mortgages.

\vspace{-0.3cm}

\subsection*{Information Retrieval Task: Relevance Analysis} % Danilo
\vspace{-0.1cm}
    The first phase consists on an explicit IR task, for which the goal is to retrieve the relevant articles that can be used to correctly answer a given yes/no question. The challenge in this task is to determine the relative relevance, i.e., Relevance Analysis (RA), of an article to the query presented in the question. Different articles dealing with the same topic often have similar wording and it is common for questions not to refer to topic keywords or refer to alternative ones. Furthermore, the restricted size of the Japanese Civil Code means that obtaining reliable linguistic information from articles is difficult and most questions will present new language structures that can range from useful to necessary for answering.
    
\vspace{-0.3cm}

\subsection*{Simple Question Answering Task: Textual Entailment} % Chien
\vspace{-0.1cm}

The goal of Textual Entailment (TE) is to decide whether a legal query/question can be answered by a set of relevant articles retrieved with RA. This task can be accomplished by recognizing textual entailment (RTE), in which the query/question is treated as an hypothesis and relevant articles as evidence. Given a question \emph{Q} and a set of relevant articles \emph{A}, ($A=\{a_1,...,a_n\}$), if \emph{Q} is answered by $a_i$ ($1\leqslant i \leqslant n$), then $a_i$ entails \emph{Q} \cite{GMDD07}, \cite{DDMR10}. A pair \emph{(Q, $a_i$)} is assigned label YES if a entailment relationship exists, i.e., $a_i$ can answer $Q$; otherwise, NO.

\vspace{-0.2cm}

\section{Proposed Approach}\label{sec:approach} % Danilo
\begin{figure*} 
        \centering
        \includegraphics[width=0.75\textwidth]{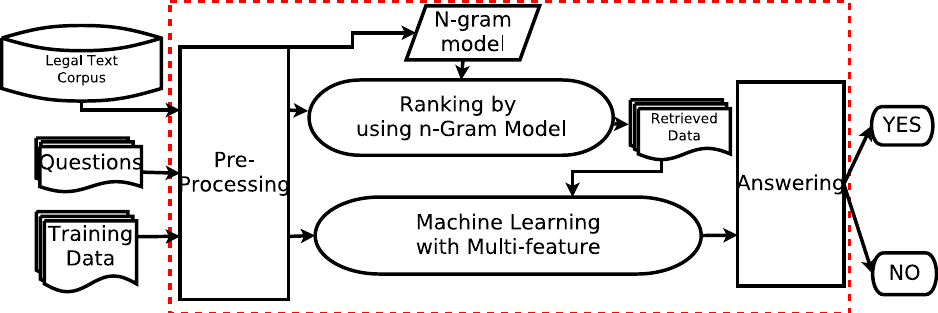}
        \caption{Model overview}\label{fig:overview}
    \end{figure*}

\vspace{-0.2cm}
    In order to be able to perform both Relevance Analysis and Textual Entailment recognition independently in phases one and two, and jointly in phase three, IR and classifier methods were developed separately. First, both the legal corpus and training data are analyzed and combined into representation models. The models are then used to rank articles or classify answers according to the task. The representation model used for Relevance Analysis is a mixed size n-gram collection and the one used for textual entailment are feature vectors for Machine Learning. Figure~\ref{fig:overview} shows the overall view of the proposed method.

    \vspace{-0.1cm}

    \vspace{-0.5cm}
	\subsection{Relevance Analysis}\label{sec:relevance-analysis} % Danilo
\begin{figure*}[h!]
	\includegraphics[width=0.9\textwidth]{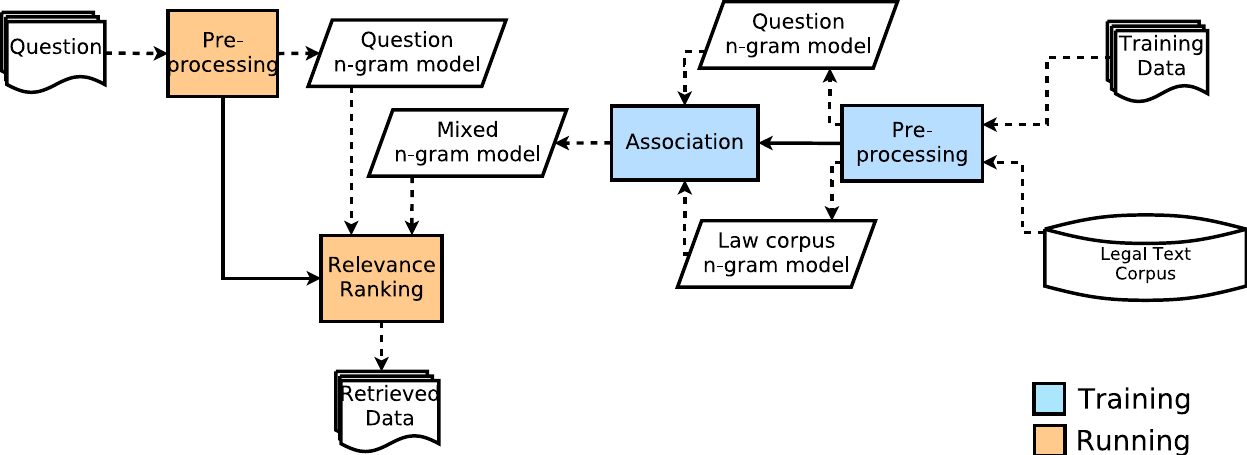}
	\caption{The process of legal text retrieval}\label{fig:IR}
\end{figure*}

    \vspace{-0.2cm}
    A detailed analysis of the Civil Code and training data revealed that lexical and syntactic overlapping may vary to a high degree between questions and articles, and also between articles concerning the same topic. However, certain morphological features, such as lemmas, retain a higher level of consistency among topics. For this reason, the adopted representation model was a mixed size n-gram model, with $n: [1, k]$, i.e., terms made by sequences up to $k$ words, in which the terms are lemmatized. For simplicity, the Relevance Analysis method hereon described was named $R_2NC$ (Ranking Related N-gram Collections). A summarized view of the process is shown on Figure~\ref{fig:IR}. 

    \noindent The steps to build the model are detailed as follows:

    \begin{enumerate}
        \item Collect the entire content for each article, including section title;
        \item Check references between articles and annotate accordingly;
        \item Tokenize and POS-tag;
        \item Remove stopwords: determiners, conjunctions, prepositions and punctuation;
        \item Lemmatize words;
        \item Generate n-grams;
        \item Expand the n-gram set, by including references n-grams;
        \item Associate article number and references;
        \item Store the model.
    \end{enumerate}

    Except for step 4, each step is responsible for adding new information to the model. The information is obtained either from the text, e.g., section title, references, or from morphological analysis, e.g., POS-tags, lemmas. If an article have references, its n-gram set is expanded with the references' n-grams. This is done so that all the necessary information for interpretation of any single article is self-contained. Besides the n-grams, links between the articles are also stored. To include the training data information, the same process is repeated for the questions, and n-gram sets from the trained questions are used to expand the associated articles' n-gram models. Since COLIEE disallowed explicit expert knowledge input, an optional information source was added after the competition, as a way of including expert knowledge in the model when available, and possibly improve system performance. This source consists in a simple term dictionary, where legal terms are associated with other correlated ones. If a given question contains n-grams referred in the dictionary, its n-gram model is expanded with the associated entries. The dictionary was written manually and contains 26 entries that were considered important after analyzing the training data, e.g., ``for a third party'' $\rightarrow$ ``to others'', and extrapolating answers to user defined queries. Tokenization and lemmatization were done using \textit{NLTK}\footnote{\href{http://www.nltk.org/}{www.nltk.org}} (v. 3.0.2) with the Punkt tokenizer and WordNetLemmatizer modules, respectively. Those modules were used with their unchanged default models and settings, trained with the Punk corpus and WordNet, respectively. POS-tagging was done using \textit{Stanford Tagger}\footnote{\href{http://nlp.stanford.edu/software/tagger.shtml}{nlp.stanford.edu/software/tagger.shtml}} (v. 3.5.2), using the unchanged \textit{english-left3words-distsim} model, which is trained on the part-of-speech tagged WSJ section of the Penn Treebank corpus.

    To determine the relative relevance of an article with regard to the content of a question, a ranking approach was adopted. First, the n-gram set of the question is obtained by applying steps 1-6, using the question content instead of article. Then, for each article in the Civil Code, a relevance score is calculated using the following formula:
    \vspace{-0.3cm}

    \begin{equation}\label{eq:relscore}
        score = \frac{\sum_{\forall t} idf(t)}{I_q \times \lvert q\_ng\_set \rvert + I_{art} \times \lvert art\_ng\_set \rvert}, \quad t \in (q\_ng\_set \cap art\_ng\_set)
    \end{equation}

    where $q\_ng\_set$ is the set of n-grams for the question, $art\_ng\_set$ is the set of n-grams for the article in the stored model, $I_q$ is the relative significance of the question n-gram set size and $I_{art}$ is the relative significance of the article n-gram set size. $idf(t)$ is the Inverse Document Frequency for the term $t$ over the articles collection
    \vspace{-0.3cm}
    
    \begin{equation}
        idf(t) = log\frac{N}{df_t}
    \end{equation}
    
    where $N$ is the total number of articles and $df_t$ is the number of articles in which $t$ appears.
    
    The formula~(\ref{eq:relscore}) is a variation of the traditional \textit{TF-IDF} scoring method, disregarding term frequency and giving different weights for the two types of document being evaluated: articles and questions, according to their size. $I_q$ and $I_{art}$ are parameters to be adjusted according to the corpus characteristics. This formula was developed during the first stages of analysis on the Civil Code corpus, when experiments with a TF-IDF based classifier showed poor results for this task and further observation showed that TF did not contribute for article relevance in many cases. As TF is absent from the formula, document size becomes a more relevant feature and must be considered in the scoring. In the studied corpus, law articles are usually much larger than questions in number of words, hence the different weights to adjust normalization of the score regarding the respective sets.

    From this point, the articles are sorted by descending score and the 10 best are selected for filtering. The filtering step consists in fetching the best scoring article and verifying if its score exceeds a parameter threshold $confidence\_thresh$. If it does, all the articles in the list that are referred by the first and exceed a parameter threshold $reference\_thresh$ are also fetched. The fetched articles compose the final list of relevant articles to the input question. Parameter adjustment is described in Section~\ref{sec:experiments}.

\vspace{-0.2cm}

\subsection{Textual Entailment}\label{sec:TE}  % Tien
\vspace{-0.1cm}

A textual entailment (TE) relation in law domain comprises two levels of information. The first level describes whether or not (YES/NO, respectively) the textual evidence addresses the hypothesis. The second level describes whether the evidence supports or opposes (YES/NO, respectively) the hypothesis. However, due to the time constraint of the competition, only the first level is explored. Therefore, semantic relations such as negation and antonym were not considered in the TE evaluation step.

To detect a TE relation on a pair $(Q, a)$, a similarity-based approach~\cite{HHNN-ICCCI-12} can be used, in which $a$ can answer $Q$ if the similarity is greater than a certain threshold. However, high level inference (see Section \ref{sec:LQA}) and the identification of the threshold make these methods more challenging to apply. We, therefore, propose to apply classification for detecting the TE relation with two advantages: (1) use of a rich feature set to represent data characteristics and (2) avoiding to identify the threshold.

This work shares most of the goals presented in Nguyen et. al.~\cite{NHNNN15}, so all the features in that work were used. However, the corpus size in this case makes it difficult to effectively train Machine Learning algorithms. For this reason, ``stronger'' features were sought as a way of compensating such problem. An additional Word2Vec feature was added to capture the semantic similarity of a pair $(Q, a)$, as observation of statistical data in Table ~\ref{tab:task2-observation} shows that the lexical overlapping may not be a strong enough feature for the classification on a $(Q, a)$ pair (e.g., cannot capture the similarity of \emph{person} and \emph{manager}). By adding the Word2Vec feature, the model aims to cover the semantic aspect instead of only lexical similarity. Word2Vec was trained by JPN Law corpus: a collection of all Civil law articles of Japan's constitution\footnote{\href{http://www.japaneselawtranslation.go.jp/}{www.japaneselawtranslation.go.jp}}. It contains 642 cleaned and tokenized articles, with about 13.5 million words in total.

For the classification, the Weka toolset~\footnote{\href{http://weka.wikispaces.com}{weka.wikispaces.com}} implementation of AdaBoost~\cite{FS97} was used, with \emph{classifier} = \textit{DecisionStump}.

\vspace{-0.4cm}

\begin{table}[ht]
\caption{Statistical data observation in phase two}
\label{tab:task2-observation}
\begin{tabular}{|l|c|c|c|c|} \hline
   & {\bf \# pairs} & {\bf \# sentences} & {\bf \# tokens} & {\bf \% uni-gram word overlapping} \\ \hline
{\bf Training Set} & 267            & 273                & 36.562          & 58.80  \\ \hline
\end{tabular}
\end{table}%\vspace{-0.5cm}

\begin{table}[ht!]
\caption{The feature groups; Avg is average; Q is a question, S is a sentence}
\label{tab:features}
\begin{tabular}{|l|l|l|} \hline
                          & \multicolumn{1}{c|}{\textbf{Feature}}    & \multicolumn{1}{c|}{\textbf{Description}} \\ \hline
\multirow{8}{*}{\textbf{Distance}} & Manhattan  & Manhattan distance from two text fragments \\ \cline{2-3} 
                          & Euclidean &  Euclidean distance from two text fragments  \\ \cline{2-3} 
                          & Cosine similarity & Cosine similarity distance \\ \cline{2-3} 
                          & Matching coefficient & Matching coefficient of two text fragments \\ \cline{2-3} 
                          & Dice coefficient & Dice coefficient of two text fragments \\ \cline{2-3} 
                          & Jaccard & Jaccard distance of two text fragments \\ \cline{2-3} 
                          & Jaro &  Jaro distance of two text fragments \\ \cline{2-3} 
                          & Damerau-Levenshtein &  Damerau Levenshtein distance of two text fragments \\ \cline{2-3} 
                          & Levenshtein     & Levenshtein distance of two text fragments \\ \hline \hline
\multirow{6}{*}{\textbf{Statistical}}  & Lcs       & The longest common sub string of two text fragments \\ \cline{2-3} 
                          & Average of TF-IDF & Term frequency-inverse document frequency \\ \cline{2-3} 
                          & Avg-TF of Q and S & Avg-TF of words in a Q appearing in a S \\ \cline{2-3} 
                          & Avg-TF of S and Q    & Avg-TF of words in a S appearing in a S  \\ \cline{2-3} 
                          & Word overlapping    & \# word overlapping in a Q appearing in a article  \\ \cline{2-3} 
                          & Average of Word2Vec & Average of word2vec similarity \\ \hline
\end{tabular}
\end{table}

\begin{figure*}
	\centering
	\includegraphics[width=0.9\textwidth]{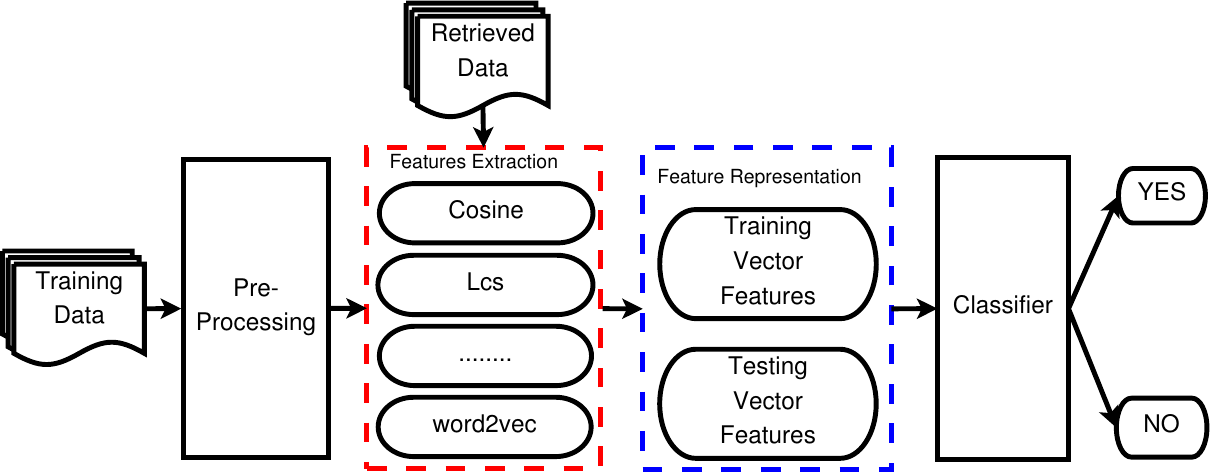}
	\caption{The process of legal textual entailment recognition}\label{fig:TE}
\end{figure*}

The features are shown in Table \ref{tab:features}, in which distance features measure distance between a question $Q$ and relevant article $a$ and statistical features capture word overlapping of this pair. After extracting features, a pipeline model was proposed and is shown in Figure \ref{fig:TE}.

In Figure \ref{fig:TE}, the first step is to preprocess the data from the input files, in which sentences and words are segmented and stopwords\footnote{https://sites.google.com/site/kevinbouge/stopwords-lists} are removed. Next, the training data is represented in a vector space model by features in Table \ref{tab:features}. The retrieved data from relevance analysis is also denoted in the same mechanism. Finally, a classifier was trained on the training data and applied on retrieved data to judge the entailment relation. Note that features in Section \ref{sec:relevance-analysis} can be also used for this task.

\section{Experiments and Results}\label{sec:experiments} % Chien
    \vspace{-0.2cm}
    \subsection{Experimental Setup}
    \vspace{-0.2cm}
    The dataset was obtained from the published data for the COLIEE shared task~\footnote{\href{http://webdocs.cs.ualberta.ca/\~miyoung2/COLIEE2015/}{webdocs.cs.ualberta.ca/\~miyoung2/COLIEE2015/}}, consisting in a text file with the Japanese Civil Code and a set of XML files with training and testing data for phases one to three. The training set for the three tasks contains 267 pairs (question, relevant articles). 
    %The testing set contains 79 questions for phases one and three and 66 questions for phase two. Phase three uses the same data as phase one.
    Experiments where divided in phases one and two only, dealing with Information Retrieval and Textual Entailment methods respectively. Each experiment comprised: i) data analysis, ii) model and parameter adjustments and iii) test runs.

    \vspace{-0.35cm}
    
    \subsection{Parameter adjustment}
    \vspace{-0.2cm}
    For $R_2NC$, parameters $I_q$, $I_{art}$, $confidence\_thresh$ (shortened to $ct$ here), \emph{reference\_thresh} ($rt$) and also $k$, the maximum n-gram size, were adjusted empirically on the training data using the following simple procedure:
    \vspace{-0.1cm}
    \begin{itemize}
        \item Starting with $I_q = 0.8$, $ct = 0.5$, $rt = 0.5$ and $k = 1$, 
        \begin{enumerate}
            \item Increase or decrease a single parameter by step $r = 0.1$ until the F-measure cannot be increased for a leave-one-out test.
            \item Repeat (1) starting from the last obtained value, with $r = 0.01$.
            \item Repeat (1) and (2) for all parameters.
        \end{enumerate}
    \end{itemize}

\begin{figure*}[ht!] 
        \centering
        \includegraphics[width=0.8\textwidth]{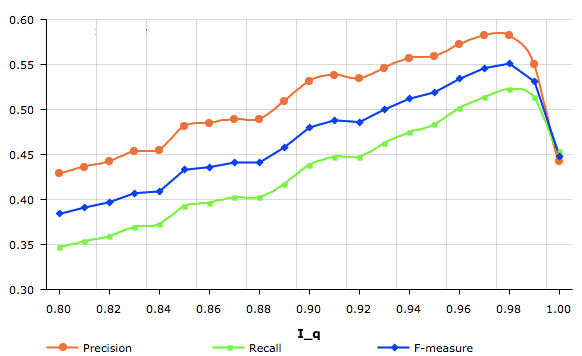} 
        \caption{Performance metrics for phase 1 related to the variation of $I_q$. $I_{art} = 1 - I_q$.} 
        \label{fig:paramadjust}
    \end{figure*}

    For $k$, step was fixed on $r = 1$. $I_q$ and $I_{art}$ respect the constraint $I_q + I_{art} = 1$. The parameters are changed in a specific order: 1. $confidence\_thesh$, 2. $k$, 3. $reference\_thresh$, 4. $I_q$. $I_q$ and $I_{art}$ respect the constraint $I_q + I_{art} = 1$
    Performance metrics were recorded for the parameter adjustment during the experiments. Fig.~\ref{fig:paramadjust} shows the performance progression on post-competition experiments for the parameters $I_q$, $I_{art}$ , with the other ones locked into their best respective values. Performance for $k \neq 3$ is negatively affected in both directions (-,+), and no further investigation was conducted for a larger range of values.

    Final parameter values used in the competition are $k = 3$, $I_q = 0.965$, $I_{art} = 0.035$, $confidence\_thresh = 0.32$ and $reference\_thresh = 0.2$.

    For the RTE classifier, default parameters from the Weka toolset\footnote{\href{http://weka.wikispaces.com}{weka.wikispaces.com}} were used for all the experiments and were not changed. The parameter values are: $iterations = 10$, $seed = 1$, no \emph{re-sampling} and $weight threshold = 100$.
    
    \vspace{-0.3cm}
	
    \subsection{Baselines}
    \vspace{-0.2cm}
    As for the second edition of COLIEE, there is still no definite baseline for the competition dataset. However, common baseline practices and related works could be used for evaluating performance on each task. For phase one, a relationship can be drawn between $R_2NC$ and TPP~\cite{LCH15}. For the TE task, the following baselines were used for comparison:
\vspace{-0.2cm}
\begin{itemize}
	\item SVMs: uses Support Vector Machines (SVMs)\footnote{https://www.csie.ntu.edu.tw/$\sim$cjlin/libsvm/} \cite{CV95} with Weka. The parameters are $C=1$, $\gamma=0$, \textit{kernel Type = radial basis function (RBF)}.
    \item AdaBoost-SVMs: uses SVMs as weak learners instead of DecisionStump.
\end{itemize}
    
    \vspace{-0.5cm}
    
 	\subsection{Evaluation Method} 
    \vspace{-0.2cm}
    Given the limited training data available, leave-one-out validation was used to evaluate the performance of the model in both tasks on the training dataset with three measures: precision (P), recall (R) and F-measure (F) as in Eq. \eqref{eq:p}, \eqref{eq:r} and \eqref{eq:f}. In phase two, accuracy (A) measurement is also used as in Eq. \eqref{eq:accuracy}.
    
    \vspace{-0.9cm}
    
   \begin{multicols}{4}
    
    \begin{equation}\label{eq:p}
		P = \frac{\mathit{Cr}}{\mathit{Rt}}
	\end{equation}
    
    \columnbreak
    \begin{equation}\label{eq:r}
		R = \frac{\mathit{Cr}}{\mathit{Rl}}
    \end{equation}
    
    \columnbreak
	
    \begin{equation}\label{eq:f}
	\hspace{-0.2cm}	
    F = \frac{2(P * R)}{P + R}
	\end{equation}
    
    \columnbreak
    
    \begin{equation}\label{eq:accuracy}
		A = \frac{\mathit{Cq}}{\mathit{Q}}
	\end{equation}
   \end{multicols}
    
   where $Cr$ counts the correctly retrieved articles for all queries, $Rt$ counts the retrieved articles for all queries, $Rl$ counts the relevant articles for all queries, $Cq$ counts the queries correctly confirmed as true or false and $Q$ counts all the queries.
    
\vspace{-0.3cm}

\subsection{Pre-competition Results}\label{sec:result}
\vspace{-0.2cm}
Pre-competition experiment results on the shared data are presented in Tables \ref{tab:task1res} and \ref{tab:task2-1-result}.

\vspace{-0.3cm}

\begin{table}[h!]
    \centering
    \caption{Experiment results for phase one (IR) with $R_2NC$. In the top 3/10 settings, articles ranked up to $3^{rd}$ or $10^{th}$ place are marked as relevant.}
    \label{tab:task1res}
    \begin{tabular}{|c|c|c|c|} \hline
        \textbf{} & \textbf{Precision} & \textbf{Recall} & \textbf{F-measure} \\ \hline
        \textbf{R2NC} & 0.568 & 0.516 & \textbf{0.54} \\ \hline
        \textbf{R2NC (top 3)} & 0.27 & 0.64 & 0.38 \\ \hline
        \textbf{R2NC (top 10)} & 0.10 & 0.77 & 0.17 \\ \hline
        \textbf{TPP (top 3)} & N/A & 0.52 & N/A \\ \hline
        \textbf{TPP (top 10)} & N/A & 0.91 & N/A \\ \hline
    \end{tabular}
 \end{table}

\vspace{-0.5cm}
    
\begin{table}
\centering
\caption{AdaBoost-DecSt (\emph{DecisionStump}) vs. SVMs  and AdaBoost-SVMs.}
\label{tab:task2-1-result}
\begin{tabular}{|p{3cm}|c|c|c|c|} \hline
   & {\bf Precision} & {\bf Recall} & {\bf F-measure} & {\bf Accuracy (\%)} \\ \hline
{\bf AdaBoost-DecSt} & 0.621 & 0.614 & 0.617 & \textbf{61.42} \\ \hline
{\bf SVMs} & 0.537 & 0.543 & 0.539 & 54.30 \\ \hline
{\bf AdaBoost-SVMs} & 0.485 & 0.491 & 0.487 & 49.06  \\ \hline
\end{tabular}
\end{table}

The results indicate that $R_2NC$ is expected to be competitive with state-of-the-art approaches to relevance analysis in legal documents, such as \textit{TPP}~\cite{LCH15}. However, the proposed method is much simpler when compared to \textit{TPP} and operates with considerably less training data: 266 documents for $R_2NC$ against 1518 documents for \textit{TPP}. $R_2NC$ design also makes it difficult for the model to be overtrained beyond the parameter adjustment, since no training data is counted more than one time and the method is single-shot, as opposed to convergence-based. Experiments were repeated with traditional TF-IDF scoring instead of $R_2NC$ formula, yielding 0.51 F-measure.

Results of RTE in Tab. \ref{tab:task2-1-result} indicate that AdaBoost with a set of appropriate features outperforms the baselines by 7.74\% (\emph{SVMs}) and 12.94\% (\emph{Adaboost-SVMs}) on F-measure. Moreover, the precision and accuracy of this method also achieve considerable improvements when compared to the baselines. This suggests that the features are expected to be efficient for addressing TE in the legal domain. This conclusion is supported by the accuracy measurements.

Another interesting point is that Word2Vec similarity contributes to improve the performance of RTE. As stated in Section~\ref{sec:LQA}, legal documents usually require concept linking to understand and answer a question; therefore, semantic similarity from Word2Vec helps to improve the performance. The results also show the efficiency of the lexical features.

The performance of RTE in the law domain, however, is not comparable with the same task in common data i.e., news articles \cite{C10,NHNNN15} due to the characteristics of law dataset, as shown in Section \ref{sec:LQA}. The performance was not improved very much even when many features in both phase one and two were combined. This suggests that more sophisticated approaches e.g., semantic inference or semantic rules should be considered in feature construction. Finally, negation and antonym analysis should be considered to improve the quality of the entailment recognition, effectively exploring the second level of entailment information as described in Section~\ref{sec:TE}.

\vspace{-0.3cm}

\subsection{Feature Evaluation}
\vspace{-0.2cm}

Further evaluation of feature impact on TE model was conducted by leave-one-out test. The most effective features are shown in Table~\ref{tab:task2-infuluential-features}.

\vspace{-0.2cm}

\begin{table}
\centering
\caption{Top 4 influential features, italic is for statistical features. Values are the difference in F-measure between the model with all features and without the single specified feature.}
\label{tab:task2-infuluential-features}
\begin{tabular}{|c|c|c|c|} \hline
{\bf Features} & {\bf Influential value} & {\bf Features} & {\bf Influential value} \\ \hline
Euclidean  & 0.005 & \emph{Lcs} & 0.0001 \\ \hline
Damerau-Levenshtein & 0.154 & Average of Word2Vec & 0.024  \\ \hline
\end{tabular}
\end{table}

Table~\ref{tab:task2-infuluential-features} shows an indication of contribution from features to the model. Results show that all effective features contribute to the method. Note that both \emph{Damerau-Levenshtein} and \emph{Euclidean} are distance features whereas \emph{the longest common substring} (lcs) is a statistical feature. 
The results support that in legal texts, there is not much word overlapping between a question and relevant articles. An interesting aspect is that Word2Vec similarity has a big positive impact to the model. This supports the conclusion on similarity stated in Section~\ref{sec:result}.

\vspace{-0.3cm}

\subsection{Competition Results}
\vspace{-0.2cm}

The method presented in this paper achieved significant results in COLIEE, being ranked 2$^{nd}$ in phase one (IR) and 3$^{rd}$ in phase three (combined IR + TE). It was not well ranked in phase two (TE). The relevant competition results are presented in Table~\ref{tab:task13compres} as they were announced in JURISIN 2015.

\vspace{-0.3cm}

\begin{table}[ht!]
    \centering
    \caption{Competition results for phases 1 (IR) and 3 (IR + TE) respectively. First three ranked.}
    \label{tab:task13compres}
    \begin{tabular}{|c|c|c|c|c|} \hline
        \textbf{Rank} & \textbf{ID} & \textbf{Prec.} & \textbf{Recall} & \textbf{F-m} \\ \hline
        1 & UA1 & 0.633 & 0.490 & 0.552 \\ \hline
        \textbf{2} & \textbf{JAIST1} & \textbf{0.566} & \textbf{0.460} & \textbf{0.508} \\ \hline
        3 & ALV2015 & 0.342 & 0.529 & 0.415 \\ \hline
    \end{tabular}
    ~
    \begin{tabular}{|c|c|c|} \hline
        \textbf{Rank} & \textbf{ID} & \textbf{Accuracy.} \\ \hline
        1 & UA1 & 0.658 \\ \hline
        2 & Kanolab3 & 0.620 \\ \hline
        \textbf{3} & \textbf{JAIST1} & \textbf{0.582} \\ \hline
    \end{tabular}
 \end{table}

 \vspace{-0.5cm}
 
 \subsection{Post-competition Analysis and Improvements}
 \vspace{-0.2cm}

Post competition analysis pointed us to possible sources of classification problems in phase 2 (TE) and also gave directions of improvement in both tasks.

For $R_2NC$, the lack of an implicit semantic mapping was an important factor when compared to the top ranked approach. To compensate for that, a term dictionary was included as a new information source for expanding the question n-gram models as described in Section~\ref{sec:relevance-analysis}. By using linguistic observations, it was possible to create basic entries in the dictionary (non-expert knowledge), improving phase 1 F-measure on the shared data (Table~\ref{tab:task1res}) from 0.54 to 0.55.

In the case of phase 2, over-fitting on training data was deemed the main factor that reduced classification performance. Our system achieved over 61\% accuracy (Table \ref{tab:task2-1-result}) when running on the shared data, but only 37.88\% reported from the competition results. Phase three results show that accuracy improved when restricting information for the classifier and this is consistent with the over-fitting assumption. Another important point is that a question $q$ and all sentences in an article $a$ were used in building the vector space model. As a result, imbalance of length between the question and the article may have affected feature calculation. This can be addressed by developing a better text segmentation method. Finally, the over-fitting assumption can also be dealt by using other classification approaches e.g., Deep Neural Networks, together with over-fitting avoidance techniques e.g., pruning, dropout.

\vspace{-0.3cm}

\subsection{Error Analysis and Discussion}
\vspace{-0.2cm}
An investigation was done on the ranked list obtained with R2NC in phase one (see Section 4.1). It revealed that relevant articles ranked 3rd and below had keywords that did not appear in the corresponding question in the corpus. This reinforces the view that the questions are highly directed, albeit in a conceptual level. Relevant articles that ranked lower than 15th (approx. 20\%) were found to require a relatively high level of abstraction to obtain an interpretation that could link to the corresponding question. Table~\ref{tab:ramiss} shows an example of complex relevance relationship.

\begin{table}[ht!]
	\scriptsize
	\caption{Example of pair (question, article) with low ranking but high relevance.}\label{tab:ramiss}
	\begin{tabular}{| c |>{}m{5.5cm}| >{}m{4cm} | c |} 
		\hline 
        \textbf{ID} & \textbf{Article} & \textbf{Question} & \textbf{Ranked in} \\
			\hline H18-2-1 & Article 697(1)A person who commences the management of a business for another person without being obligated to do so (hereinafter in this Chapter referred to as "Manager") must manage that business (hereinafter referred to as "Management of Business") in accordance with the nature of the business, using the method that best conforms to the interests of that another person (the principal).(2)The Manager must engage in Management of Business in accordance with the intentions of the principal if the Manager knows, or is able to conjecture that intention.
			& In cases where a person plans to prevent crime in their own house by fixing the fence of a neighboring house, that person is found as having intent towards the other person. &  424th \\
		\hline 
	\end{tabular}
\end{table}

\begin{table}
	\scriptsize
	\caption{Examples of entailment judgment; P is predicted and A is annotated}\label{tab:exp-judgement}
	\begin{tabular}{| c |>{}m{5.5cm}| >{}m{4cm} | c  | c |} 
		\hline \textbf{ID} & \textbf{Article} & \textbf{Question} & \textbf{P} & \textbf{A} \\
		\hline H18-2-4 & (Managers' Claims for Reimbursement of Costs)Article 702(1)If a Manager has incurred useful expenses for a principal, the Manager may claim reimbursement of those costs from the principal.(2)The provisions of Paragraph 2 of Article 650 shall apply mutatis mutandis to cases where a Manager has incurred useful obligations on behalf of the principal.(3)If a Manager has engaged in the Management of Business against the intention of the principal, the provisions of the preceding two paragraphs shall apply mutatis mutandis, solely to the extent the principal is actually enriched.
			& In cases where a person repairs the fence of a neighboring house after it collapsed due to a typhoon, but the neighbor had intended to replace the fence with a concrete-block wall in the near future, if a separate typhoon causes the repaired sections to collapse the following week, reimbursement of repair fees can no longer be demanded. & YES & YES \\
			\hline H18-26-1 & (Renunciation of Shares and Death of Co-owners)Article 255
If one of co-owners renounces his/her share or dies without an heir, his/her share shall vest in other co-owners.
			& In cases where person A and person B co-own building X at a ratio of 1:1, if person A dies and had no heirs or persons with special connection, ownership of building X belongs to person B. & NO & YES \\
		\hline 
	\end{tabular}
\end{table}

Table \ref{tab:exp-judgement} shows a case in which our system gives correct outputs (ID H18-2-4). In this example, there are several common words from which this approach can correctly judge the TE relation, e.g., \emph{reimbursement}. In addition, several words can be inferred from the questions by using Word2Vec similarity e.g., \emph{person $\sim$ manager}, \emph{fees $\sim$ costs or expenses}. This supports our observation that TE can be addressed by using lexical features and word similarity. For example, in (ID H18-2-4), our system can still predict the TE relation correctly, even with little lexical overlap. This indicates the efficiency of this approach, and especially of the word similarity feature.

On the other hand, the pair H18-26-1 exemplifies a case in which the system predicted NO while TE relation was annotated YES even when the question and answer share more common words. This shows the limitation of this feature set in cases where the question and answer are short. In this case, after removing stop words, a few remaining words may not be enough to capture the TE relation. Moreover, the lack of important words e.g., \emph{building}, \emph{connection} or \emph{belong} reveals a big challenge for our system to decide the TE relation. This suggests that a keyword enriching mechanism such as term expansion used in phase one could improve the results.

In order to facilitate the understanding of different error cases and give other people the opportunity to try the system developed for the competition, an online demo system\footnote{\href{http://150.65.242.101:3001/}{http://150.65.242.101:3001/}} has been made available. In this demo it is possible to input user defined questions or just verify the answers to questions in the COLIEE shared data.

\section{Conclusion}\label{sec:conclusion}
This paper explores the challenging issue of building a QA system in the legal domain. We propose a model including three stages: legal information retrieval, legal textual entailment and legal text answering.
In the first stage, a mixed size n-gram model built from morphological analysis is used to rank and select relevant articles corresponding to a legal question; next, pairs of questions and retrieved articles are judged by a machine learning algorithm trained on lexical features and Distributional Semantic similarity, to decide whether the questions can be answered positively or negatively by the retrieved articles; and finally, correct answers would be provided for users in the final stage. The contributions of this work in IR and TE task are: 1) a simple, yet effective language model for law corpora coupled with a Relevance Analysis method ($R_2NC$) capable of exploiting such model; 2) The use of TF-IDF and Word2Vec similarity features for applying Machine Learning algorithms to RTE. With a recall of 0.64 for the top 3 ranked articles, $R_2NC$ appears as competitive when compared to state-of-the-art similar work, in spite of being more simple and applicable with less training data. By combining lexical features and Word2Vec similarity, this approach for LQA also outperformed the baselines by 8.4\% (\emph{SVM}) and 11.3\% (\emph{Adaboost-SVMs}) on F-measure. Results in the COLIEE competition for the IR task (0.508 F-measure, 2$^{nd}$ place) and the combined IR+TE task (0.582 accuracy, 3$^{rd}$ place) indicate a substantial adequacy to the tasks addressed. The competition also provided important shortcomings of the proposed approach, namely the lack of implicit semantic representation and classifier over-fitting. Those shall be addressed in future work.

Still on future directions, information on a higher abstraction level, e.g., syntactic mappings, could be used to improve the language model for the IR task. In the TE task, since a sentence in a legal article is usually long, a sophisticated method of sentence partition e.g., requisite and effectuation should be considered. In feature extraction, features in IR should be combined with lexical features in TE and investigated to improve the quality of the judgment. Moreover, capturing contradictions in the TE relation by current statistical features is a big challenge. To solve this issue, semantic rules over negation and antonym detection should be defined and incorporated into the feature extraction. Finally, we would like to investigate and apply sentence similarity calculation by Sent2Vec to improve the performance of the TE.

\vspace{-0.3cm}

\section*{Acknowledgements}
\vspace{-0.2cm}
This work is supported partly by the grant of NII Research Cooperation and JAIST's Research grant.
\vspace{-0.2cm}
%
% ---- Bibliography ----
%
\begin{scriptsize}

\end{scriptsize}

\end{document}